\documentclass[%
reprint,
showpacs,
 amsmath,amssymb,
 aps,
]{revtex4-2}

\usepackage[utf8]{inputenc}

\usepackage{graphicx}
\usepackage{dcolumn}
\usepackage{bm}
\usepackage{mathrsfs} 
\usepackage{physics}
\usepackage[dvipsnames]{xcolor}

\usepackage{float}

\usepackage[]{hyperref}
\usepackage{tikz}
\usetikzlibrary{arrows,shapes}
\usetikzlibrary{trees}
\usetikzlibrary{matrix,arrows} 				
\usetikzlibrary{positioning}				
\usetikzlibrary{calc,through}				
\usetikzlibrary{decorations.pathreplacing}  
\usepackage{pgffor}							
\usetikzlibrary{decorations.pathmorphing}	
\usetikzlibrary{decorations.markings}
\tikzset{
	vector/.style={decorate, decoration={snake}, draw},
	provector/.style={decorate, decoration={snake,amplitude=2.5pt}, draw},
	antivector/.style={decorate, decoration={snake,amplitude=-2.5pt}, draw},
	fermion/.style={draw=black, postaction={decorate},
	decoration={markings,mark=at position .55 with {\arrow[draw=black]{>}}}},
	fermionbar/.style={draw=black, postaction={decorate},
	decoration={markings,mark=at position .55 with {\arrow[draw=black]{<}}}},
	fermionnoarrow/.style={draw=black},
	gluon/.style={decorate, draw=black,
	decoration={coil,amplitude=4pt, segment length=5pt}},
	scalar/.style={dashed,draw=black, postaction={decorate},
	decoration={markings,mark=at position .55 with {\arrow[draw=black]{>}}}},
	scalarbar/.style={dashed,draw=black, postaction={decorate},
	decoration={markings,mark=at position .55 with {\arrow[draw=black]{<}}}},
	scalarnoarrow/.style={dashed,draw=black},
	electron/.style={draw=black, postaction={decorate},
	decoration={markings,mark=at position .55 with {\arrow[draw=black]{>}}}},
	bigvector/.style={decorate, decoration={snake,amplitude=4pt}, draw},
}

\def\dm{\Delta m^2}

\newcommand{\mmean}[1]{\overline{m^2_{#1}}}

\newcommand{\Lagr}{\mathscr{L}}

\definecolor{DiracColor}{rgb}{0.0,0.55,0.0}
\colorlet{MajoranaColor}{blue}

\begin{document}


\title{
	   Absolute neutrino mass and the Dirac/Majorana distinction\\
	   from the weak interaction of aggregate matter
}

\author{Alejandro Segarra}%
\author{Jos\'e Bernab\'eu}
\affiliation{%
	Department of Theoretical Physics, University of Valencia
	and IFIC, Univ.~Valencia - CSIC, Burjassot, Valencia, Spain
}%

\date{\today}

\begin{abstract}
The 2$\nu$-mediated force has a range of microns, well beyond the atomic scale. 
The effective potential is built from the $t$-channel absorptive part of the scattering amplitude 
and depends on neutrino properties on-shell. 
We demonstrate that neutral aggregate matter has a weak charge 
and calculate the matrix of six coherent charges for its interaction with definite-mass neutrinos. 
Near the range of the potential the neutrino pair is non-relativistic, 
leading to observable absolute mass and Dirac/Majorana distinction 
via different $r$-dependence and violation of the weak equivalence principle.
\end{abstract}

\maketitle


The experimental evidence of neutrino oscillations 
is one of the most important discoveries in particle physics.
First evidence of model-independent neutrino oscillations
were obtained in 1998 by the Super-Kamiokande atmospheric neutrino experiment~\cite{Fukuda:1998mi}, 
in 2002 by the SNO solar neutrino experiment~\cite{Ahmad:2002jz}, 
and later with reactor~\cite{Araki:2004mb} and accelerator~\cite{Ahn:2006zza} neutrinos.
The existence of neutrino oscillations implies that neutrinos are massive particles 
and that the three flavor neutrinos $\nu_e,\, \nu_\mu,\, \nu_\tau$
are mixtures of the neutrinos with definite masses $\nu_i$ (with $i = 1,\, 2,\, 3$). 
The phenomenon of neutrino oscillations is being studied in a variety of experiments
which fully confirm this fundamentally quantum phenomenon 
in different disappearance and appearance channels. 
The mixing matrix $U_\mathrm{PMNS}$~\cite{Pontecorvo:1957cp, Maki:1962mu} contains three mixing angles, 
already known, 
and one CP-violating phase for flavor oscillations. 
This quantum interference phenomenon measures the phase differences 
due to the squared mass splittings $\Delta m^2_{21}$ and $\abs{\Delta m^2_{31}}$, 
but the absolute mass scale is inaccessible. 
The answer to this last open question 
is being sought after by the KATRIN experiment~\cite{Aker:2019uuj}
in tritium beta decay, with an expected final sensitivity around $0.2$~eV.

Knowing that neutrinos are massive, 
the most fundamental problem is the determination of the nature of neutrinos with definite mass: 
are they either four-component Dirac particles with a conserved global lepton number $L$, 
distinguishing neutrinos from antineutrinos, 
or two-component truly neutral (no electric charge and no global lepton number) 
self-conjugate Majorana particles~\cite{Majorana:1937vz}? 
For Dirac neutrinos, like quarks and charged leptons, 
their masses can be generated in the Standard Model of particle physics 
by spontaneous breaking of the gauge symmetry with the doublet Higgs scalar, 
if there were additional right-handed sterile neutrinos. 
But the Yukawa couplings would then be unnaturally small compared to those of all other fermions. 
One would also have to explain the origin of the global lepton number 
avoiding a Majorana mass for these sterile neutrinos. 
A Majorana $\Delta L = 2$ mass term, with the active left-handed neutrinos only, 
leads to definite-mass neutrinos with no definite lepton charge. 
However, there is no way in the Standard Model to generate this Majorana mass, 
so the important conclusion in fundamental physics arises: 
Majorana neutrinos would be an irrefutable proof of physics beyond the Standard Model. 
Due to the Majorana condition of neutrinos with definite mass being their own antiparticles, 
Majorana neutrinos have two additional CP-violating phases~\cite{Bilenky:1980cx, Doi:1980yb, Bernabeu:1982vi} 
beyond the Dirac case.
  
Neutrino flavor oscillation experiments cannot determine
the fundamental nature of massive neutrinos. 
In order to probe whether neutrinos are Dirac or Majorana particles, 
the known way has been the search of processes violating the global lepton number $L$. 
The difficulty encountered in these studies is well illustrated by the so-called 
\emph{confusion theorem}~\cite{Case:1957zza, ryan1964equivalence}, 
stating that in the limit of zero mass there is no difference between Dirac and Majorana neutrinos. 
As all known neutrino sources produce highly relativistic neutrinos 
(except for the present cosmic neutrino background in the universe), 
the $\Delta L = 2$ observables are highly suppressed. 
Up to now, there is a consensus that the highest sensitivity to small Majorana neutrino masses 
can be reached in experiments on the search of the $L$-violating 
neutrinoless double-$\beta$ decay process ($0\nu\beta\beta$). 
Dozens of experiments around the world are seeking out a positive signal,
and the most sensitive limits are obtained by GERDA-II~\cite{Agostini:2018tnm} in $^{76}$Ge, 
CUORE~\cite{Adams:2019jhp} in $^{130}$Te 
and KAMLAND-Zen~\cite{KamLAND-Zen:2016pfg} in $^{136}$Xe. 
An alternative to $0\nu\beta\beta$ is provided by 
the mechanism of neutrinoless double electron capture ($0\nu$ECEC)~\cite{Bernabeu:1983yb}, 
which actually corresponds to a virtual mixing 
between a nominally stable parent $Z$ atom and a daughter $(Z-2)^*$ atom with two electron holes. 
The experimental process is the subsequent X-ray emission 
and it becomes resonantly enhanced when the two mixed atomic states are nearly degenerate. 
The search of appropriate candidates~\cite{Eliseev:2011zza} satisfying the resonant enhancement 
is being pursued with the precision in the measurement of atomic masses provided by traps. 
The process can be stimulated~\cite{Bernabeu:2017ape} in XLaser facilities. 
The $2\nu$ECEC decay, allowed in the Standard Model, 
has recently been observed for the first time by the XENON Collaboration~\cite{XENON:2019dti} in $^{124}$Xe. 
This last process, contrary to the case of $2\nu\beta\beta$ for searches of $0\nu\beta\beta$, 
is not an irreducible background for $0\nu$ECEC when the resonance condition is satisfied.

In this paper we present and develop a novel idea for this dilemma, 
following a different path to the search of $\Delta L = 2$ processes. 
It is based on having a pair of virtual non-relativistic neutrinos of definite mass, 
whose quantum distinguishability is different for Dirac and Majorana nature due to the lepton charge. 
Such a physical situation is apparent in the \emph{long-range} force mediated by two neutrinos at distances near its range.
There is a conjunction of facts that cooperate in the achievement of this goal:

\begin{enumerate}

	\item 
		The Compton wavelength of massive neutrinos is of order 1 micron.
		Although the absolute scale of neutrino masses is still unknown,
		the present upper limit and the known $\abs{\Delta m^2_{31}}$ and $\Delta m^2_{21}$	values 
		---see, for example, Ref.~\cite{deSalas:2017kay}--- 
		tell us that $m_\nu \sim 0.1$~eV can be taken as a reference. 
		Such a range for the two-neutrino-mediated force is well above the atomic scale, 
		so the force will be operative for atoms and aggregate matter if they have a weak charge, 
		being neutral in electric charge.
 
	\item
		Indeed a coherent weak charge~\cite{Feinberg:1989ps, Bernabeu:1991xj, Hsu:1992tg} is built from 
		neutral-current interactions of neutrinos with electrons, protons and neutrons,
		and the charged-current interaction of electron neutrinos with electrons. 
		These different weak charges for $\nu_e$ and $\nu_{\mu,\tau}$ are
		proportional to the number operator and thus they violate the weak equivalence principle (WEP).

	\item
		For neutral-current interactions, 
		flavor mixing is unoperative and the intermediate neutrino propagation with definite mass directly appears. 
		For the charged-current interaction, 
		the mixing $U_{ei}$ of electron neutrinos to all neutrinos of definite mass will be needed. 
		This ingredient is also well known~\cite{deSalas:2017kay} from neutrino oscillation experiments.
 
	\item
		The dispersion theory of long-range forces leads to the effective potential 
		in terms of the absorptive part of the amplitude at low $t$, 
		i.e. the energy of the neutrino pair in the $t$ channel.
		Hence the physics involved, by unitarity, is that of a pair of low energy neutrinos with definite mass. 
		One then expects a hermitian matrix with six different weak charges. 
		The only unknown is the lightest neutrino mass. 

	\item
		For Dirac neutrinos with definite lepton charge, 
		the interaction vertex is the chiral charge distinguishing neutrinos from antineutrinos. 
		For Majorana neutrinos with no conserved charge, 
		the interaction vertex is twice the axial charge and so, contrary to the Dirac case, the pair is in P wave. 
		The absorptive parts for Dirac and Majorana neutrinos will differ in the mass-dependent terms
		leading to different $r$-dependent potentials near their range.

	\item
		Formidable precision tests of WEP 
		and the $r$ dependence of forces between matter aggregates
		are being pursued in recent years. 
		They reached the centimeter to micron scale of distances from different approaches like 
		torsion balance~\cite{Kapner:2006si}, 
		optical levitation~\cite{Rider:2016xaq}
		and
		atom interferometry~\cite{Jaffe:2016fsh, Sabulsky:2018jma}.

\end{enumerate}

Following these guiding concepts, 
we have developed~\cite{Segarra:2019bee} the theory of the long-range force mediated by two neutrinos 
including for the first time all ingredients of neutrino physics 
relevant to the region of distances near the range of the interaction, 
with masses, mixing and the Dirac/Majorana distinction. 

The dispersion theory of long-range forces was developed in Ref.~\cite{Feinberg:1970zz}
for the two-photon mediation between neutral objects, 
reproducing in a model-independent way the Casimir-Polder potential~\cite{Casmir:1947hx}. 
The method was later applied to the case of charged-neutral objects~\cite{Bernabeu:1976jq}
and extended for a low-momentum-transfer theorem in lepton-hadron scattering~\cite{Penarrocha:1980fx}. 
The two-neutrino mediation was given in 1968~\cite{Feinberg:1968zz} 
for charged-current interactions of two electrons and, 
in the dispersion approach, it was later extended~\cite{Feinberg:1989ps, Bernabeu:1991xj, Hsu:1992tg}
considering neutral current interactions as well. 
With the advent of neutrino masses and mixings, 
the long range $r^{-5}$ potential will be modified at distances near its range. 
For Dirac neutrinos, these effects have been calculated~\cite{Thien:2019ayp} for electrons and nucleons 
using old-fashioned perturbation theory. 
In this work we present the results for the effective potential between aggregate matter 
obtained in the dispersion approach, 
with appropriate treatment of the different mass terms corresponding to either Dirac or Majorana neutrinos. 
Thus we open an alternative path to the known search of forbidden $\Delta L = 2$ processes. 
We demonstrate that, at distances 1--10 microns, 
the potential is extremely sensitive to the mass of the lightest neutrino varying between 0 and $0.1$~eV 
and to the Dirac/Majorana distinction.

\begin{figure}[b]
	\begin{tikzpicture}[line width=1 pt, scale=1.2]
		\draw[fermion](150:1.5) -- (150:0.3cm);
			\node at (150:1.9) {$A$};
		\begin{scope}[shift={(35:4pt)}]
			\draw[fermion](150:1.5) -- (150:0.3cm);
		\end{scope}
		\begin{scope}[shift={(35:-4pt)}]
			\draw[fermion](150:1.5) -- (150:0.3cm);
			\draw (150:0.3) -- (150:0);
		\end{scope}
		
		\draw[fermionbar](30:1.5) -- (30:0.3cm);
			\node at (30:1.9) {$A$};
		\begin{scope}[shift={(-35:-4pt)}]
			\draw[fermionbar](30:1.5) -- (30:0.3cm);
		\end{scope}
		\begin{scope}[shift={(-35:4pt)}]
			\draw[fermionbar](30:1.5) -- (30:0.3cm);
			\draw (30:0.3) -- (30:0);
		\end{scope}
		
		\draw[fill=black] (0,0) circle (.3cm);
		\draw[fill=white] (0,0) circle (.29cm);
		\begin{scope}
		    	\clip (0,0) circle (.3cm);
		    	\foreach \x in {-.9,-.8,...,.3}
				\draw[line width=1 pt] (\x,-.3) -- (\x+.6,.3);
	  	\end{scope}
	  	
	  	\begin{scope}
	  		\clip (-2,-2) rectangle (0,0);
		  	\draw[fermion] (0,-0.9) ellipse (0.3 and 0.6);
		  	\node at (-0.5, -0.75) {$\nu$};
	  	\end{scope}
	  	\begin{scope}
	  		\clip (2,-2) rectangle (0,0);
		  	\draw[fermion] (0,-0.9) ellipse (-0.3 and -0.6);
		  	\node at (0.5, -1.05) {$\nu$};
	  	\end{scope}
	  	
	  	\draw[dashed, thin] (-2,-0.9) -- (2, -0.9);
	  	\node at (2.2, -0.9) [above left] {\footnotesize t channel};
			
	  	\begin{scope}[shift={(0,-1.8)}]
			\draw[fermion](-150:1.5) -- (-150:0.3cm);
				\node at (-150:1.9) {$B$};
			\begin{scope}[shift={(-35:4pt)}]
				\draw[fermion](-150:1.5) -- (-150:0.3cm);
			\end{scope}
			\begin{scope}[shift={(-35:-4pt)}]
				\draw[fermion](-150:1.5) -- (-150:0.3cm);
				\draw (-150:0.3) -- (-150:0);
			\end{scope}
			
			\draw[fermionbar](-30:1.5) -- (-30:0.3cm);
				\node at (-30:1.9) {$B$};
			\begin{scope}[shift={(35:-4pt)}]
				\draw[fermionbar](-30:1.5) -- (-30:0.3cm);
			\end{scope}
			\begin{scope}[shift={(35:4pt)}]
				\draw[fermionbar](-30:1.5) -- (-30:0.3cm);
				\draw (-30:0.3) -- (-30:0);
			\end{scope}
			
			\draw[fill=black] (0,0) circle (.3cm);
			\draw[fill=white] (0,0) circle (.29cm);
			\begin{scope}
			    	\clip (0,0) circle (.3cm);
			    	\foreach \x in {-.9,-.8,...,.3}
					\draw[line width=1 pt] (\x,-.3) -- (\x+.6,.3);
			\end{scope}
		\end{scope}
	\end{tikzpicture}

	\caption{
	 	Leading-order Feynman diagram for the neutrino-pair mediated 
		$t$-channel $AB\to AB$ scattering.
	}
	\label{fig:ABABt}
\end{figure}
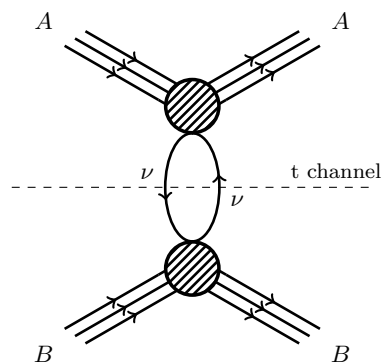

The Feynman diagram describing the two-neutrino-mediated interaction 
between two objects A and B is depicted in Figure~\ref{fig:ABABt}.
In the dispersion approach the potential is given by the integral transform of 
the $t$-channel absorptive part of the scattering amplitude  $A B \to A B$. 
For three neutrino species cut on-shell, there are six different amplitudes and thresholds 
$t_{ij} = (m_i + m_j)^2$,
such that 
\begin{equation}
	V(r) = \frac{-1}{4\pi^2 r}\, 
	\sum_{ij} \int_{t_{ij}}^\infty \, \dd t\, 
	e^{-\sqrt{t} \, r}\,
	\mathrm{Im}~{\cal M}_{ij}(t) \,.
	\label{eq:transform}
\end{equation}

\noindent
We already appreciate in Eq.~(\ref{eq:transform}) the complementarity, 
in the quantum-mechanical sense, 
between long distances $r$ and low $t$ behavior of the absorptive parts. 
Using Unitarity, 
they are given by the $t$-channel crossing for $A \bar A \to B \bar B$~\cite{martin1970elementary}
\begin{equation}
	\begin{aligned} 
		 &\mathrm{Im}~\mathcal{M}_{ij}^{A\bar A \to B\bar B} =
		\frac{1}{2} \int \frac{\dd^4 k_1}{(2\pi)^3}\, \delta(k_1^2 - m_i^2)\,
		\frac{\dd^4 k_2}{(2\pi)^3}\, \delta(k_2^2 - m_j^2)
		\\
		&\hspace{0cm} \times
		(2\pi)^4 \delta^{(4)}(k_1+k_2-p_i) 
		{\cal M}_{B\bar B\to\nu_i \bar \nu_j}^*{\cal M}_{A\bar A\to\nu_i \bar\nu_j}^{}
		\,.
	\end{aligned}
	\raisetag{-0.125cm}
	\label{eq:ImMABAB}
\end{equation}

\noindent
As seen, the intermediate states are pairs ($i,\, j$) 
of definite-mass neutrinos connected to their interaction vertices with matter. 
For neutral-current interactions there are diagonal $i = j$ terms only, 
interacting with electrons, protons and neutrons. 
For charged-current interactions there are non-diagonal terms $i \neq j$ too for electron neutrinos ($\alpha = e$)
interacting with electrons by means of the mixing $U_{ei}^{} U_{ej}^*$.

The low-energy interaction of definite-mass neutrinos with the matter constituents 
is given by the effective Lagrangian in the Standard Model
\begin{equation}
	\Lagr_\mathrm{eff} = -\frac{G_F}{2\sqrt{2}}\,
	\left[ \bar \nu_j\,  \gamma^\mu\,(1-\gamma_5)\, \nu_i \right] 
	\left[ \bar \psi \,  \gamma_\mu\,(g^\psi_{V_{ij}} - g^\psi_{A_{ij}} \gamma_5)\, \psi \right]\,,
	\label{eq:effLagrij}
\end{equation}

\noindent
with $\psi = e,\, p,\, n$ describing the fermion fields of the constituents.
For aggregate matter, 
we are interested in the current component proportional to the number operator, 
the $\mu = 0$ component of the vector current. 
Thus the only relevant couplings are
\begin{align}
	\nonumber
	g_{V_{ij}}^{e} &= 2 U_{ei}^{} U_{ej}^* - ( 1 - 4 \sin^2\theta_W^{} ) \delta_{ij}\,, \\
	\nonumber
	g_{V_{ij}}^{p} &= ( 1 - 4 \sin^2\theta_W^{}) \delta_{ij}\,,  \\ 
	g_{V_{ij}}^{n} &= - \delta_{ij}\,,
\end{align}

\noindent
and the coherent global weak charges are given by the six independent elements of the hermitian matrix
\begin{equation}
	Q^{ij}_W (Z, N) = Z (g_{V_{ij}}^e + g_{V_{ij}}^p) + N g_{V_{ij}}^{n}\,,
	\label{eq:Qij}
\end{equation}

\noindent
where both the atomic number $Z$ and the neutron number $N$ have to be multiplied 
by the number of neutral atoms in the A (or B) object.
As gravity, we have here a coherent interaction for aggregate matter, 
however not proportional to the mass of the object, 
thus leading to a violation of WEP.

In the calculation of the absorptive part in Eq.~(\ref{eq:ImMABAB})
there are both dynamical and kinematical mass effects.
For Dirac neutrinos, 
the right-handed components of the states with definite mass are sterile, 
so their interaction continues to be the V-A chiral charge. 
Majorana neutrino states of definite mass,
on the other hand,
have their two chiralities of left-handed neutrino and its conjugate as active interacting components. 
This is the case because
of the $\Delta L = 2$ Majorana mass term in the Langrangian,
connecting the left-handed field and its conjugate. 
As a consequence, 
the interaction vertex of this state is twice the axial current and, 
when contracted with the coherent weak charge of matter, 
the intervening contribution is the parity-odd axial charge. 
This fundamental difference in the dynamics of Majorana neutrinos, with respect to Dirac neutrinos, 
is affecting the behavior of the absorptive part at low values of $t$. 
The results for either Dirac or Majorana absorptive parts are
\begin{align}
	\nonumber
	&\text{Im}~{\cal M}_{ij} = -\frac{G_F^2}{48\pi}\,
	t  \,
	Q_{W, A}^{ij} Q_{W, B}^{ij\;\,*} 
	\sqrt{ 1 - \frac{4\mmean{ij}\,}{t} + \left[ \frac{\dm_{ij}}{t} \right]^2 }
	\\
	&\; \hspace{0.25cm}\times 
	\left[ 1 - \frac{1}{t} \left\{ \begin{aligned}
			&\textcolor{DiracColor}{\mmean{ij}}\\
			&\textcolor{MajoranaColor}{\mmean{ij} + 3 m_i m_j}
		\end{aligned}\right\}
	- \frac{1}{2}\left[ \frac{\dm_{ij}}{t} \right]^2  \right]
	\,,
	\label{eq:ImM}
\end{align}
where the \textcolor{DiracColor}{upper}/\textcolor{MajoranaColor}{lower} dynamical terms in the bracket 
correspond to \textcolor{DiracColor}{Dirac}/\textcolor{MajoranaColor}{Majorana} neutrinos,
whereas mass effects in the first line are kinematical 
and thus blind to the neutrino nature.

Several relevant comments are in order:
(i) the charges 
are now specific to the ($i,\, j$) intermediate channels
as $Q_W^{ij} = 2Z U_{ei} U_{ej}^* - N \delta_{ij}$ in Eq.~(\ref{eq:Qij});
(ii) the complex mixings enter the absorptive part, 
although only as the $\abs{U_{ei}}$ moduli even for $A \neq B$ matter aggregates,
so that no CP violation effects are accessible; 
(iii) Eq.~(\ref{eq:ImM}) leads in the limit of vanishing masses to the correct linear t dependence; 
(iv) the Dirac/Majorana distinction appears in the mass terms, 
and they reproduce the case of a single neutrino species~\cite{Grifols:1996fk} 
considered in the context of neutron stars;
(v) for non-relativistic neutrinos with momenta $k$ in the CM frame 
$t = t_0 (1 + k^2/ m_i m_j)$, 
so the absorptive part of the ($i,\, j$) channel is 
either S-wave proportional to $k$ for Dirac neutrinos
or P-wave proportional to $k^3$ for Majorana neutrinos.
This different non-relativistic behavior leads to the final distinction in the potential,
providing the signal to determine the neutrino nature.

\begin{figure}
	\centering
	\begin{tikzpicture}[line width=1 pt, scale=1]
		\node at (0,0){
			\includegraphics[width=0.45\textwidth]{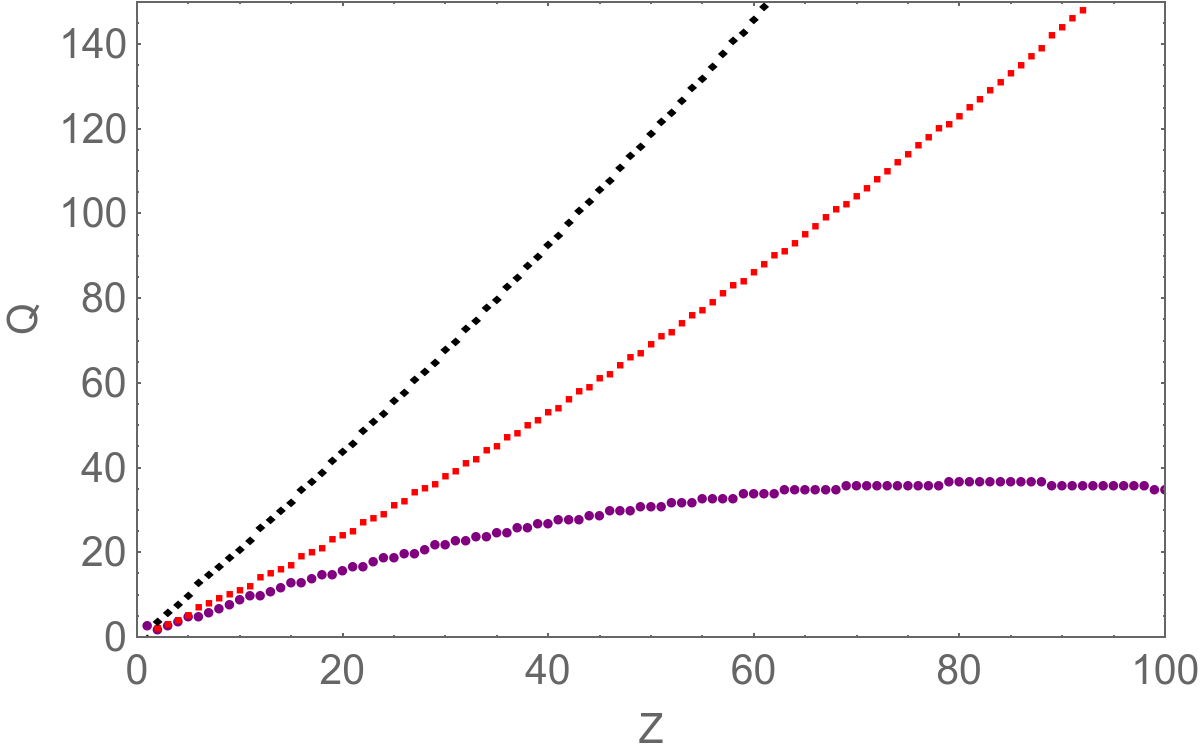}
		};

			\node at (3,-0.4){\textcolor{Purple}{$Q_W^e$}};
			\node at (3,1.6){\textcolor{Red}{$-Q_W^{\mu, \tau}$}};
			\node[rotate=48] at (0.3,2.0){Gravit.};
	\end{tikzpicture}
	\caption[Weak flavor charges]{
		$e$ (\tikz{\draw[Purple, fill=Purple] (0,0) circle (.4ex);}),
		$\mu$ and $\tau$ (\tikz{\fill [Red] (0.1,0.1) rectangle (0.2,0.2);})
		weak flavor charges
		of the elements with $(Z,N)$ in the valley of stability,
		as well as their gravitational coupling, approximately proportional to $Z+N$ 
		(\tikz{\fill [black, rotate=45] (0.1,0.1) rectangle (0.2,0.2);}).
		Beware a minus sign in the $\mu, \tau$ flavor charges.
	}
	\label{fig:qweak}
\end{figure}

If the lightest neutrino is massless, 
its contribution will dominate the behavior of the potential at the longest distances
with the known $r^{-5}$ dependence.
At distances such that  $m_i \, r \ll 1\; \forall i$, 
the exchanged neutrinos are extremely relativistic, 
thus leading to a common absorptive part linear in $t$. 
In this last case,
there is a global A-B coupling of coherent weak flavor charges $Q_W^\alpha$
\begin{equation}
	\sum_{ij} Q_{W,A}^{ij} (Q_{W,B}^{ij})^* = 
	\sum_{\alpha} Q_{W,A}^{\alpha}\, Q_{W,B}^{\alpha}\,,
	\label{eq:globalcharges}
\end{equation}
as a sum extended to all diagonal neutrino $\alpha$-flavors, 
with $Q_W^e = 2Z - N$ and $Q_W^{\mu,\tau} = - N$ for each single atom. 
These weak flavor charges are represented in Figure~\ref{fig:qweak} for the most stable isotopes, 
following a semi-empirical formula~\cite{krane1987introductory} relating $Z$ and $N$. 
The $Z$ dependencies of the weak flavor charges are compared with the $Z + N(Z)$ curve 
approximately giving the mass coupling for gravity.
In this short-distance limit, 
using Eq.~(\ref{eq:globalcharges}) for aggregate matter reproduces the known repulsive potential
\begin{equation}
	V (m_i\, r \ll 1) = \frac{G_F^2}{16 \pi^3}\,
	\frac{1}{r^5}\,
	\sum_{\alpha} Q_{W,A}^{\alpha}\, Q_{W,B}^{\alpha}\,.
	\label{eq:Vshort}
\end{equation}

Having identified the physics responsible of the Dirac/Majorana neutrino nature distinction
in our interaction potential between two objects of ordinary matter, 
we find it convenient to show explicit analytic results for the $r$ dependence near its range. 
Expanding Eq.~(\ref{eq:ImM}) and inserting it in the integral transform (\ref{eq:transform}),
we obtain the contribution of each ($i,\, j$) intermediate channel,
\begin{align}
	\nonumber
	&V ( m_i\, r > 1)
	= \frac{G_F^2}{64 \pi^{5/2}}\,
	\sum_{ij} Q_{W,A}^{ij} (Q_{W,B}^{ij})^* \,
	e^{-r\sqrt{t^{}_{ij}}}
	\\
	  &\hspace{0cm} \times
	\frac{\sqrt{t^{}_{ij}}}{r} \,
	\left[ \frac{ 2\mu^{}_{ij} }{ r} \right]^{3/2}
	\left[
		\left\{
			\begin{aligned}
				\textcolor{DiracColor}{1}\\\textcolor{MajoranaColor}{0}
			\end{aligned}
		\right\}
		+ \left( 3- \frac{4 \mu^{}_{ij}}{\sqrt{t^{}_{ij}}} \right) \frac{1}{2 \mu^{}_{ij} r}
	\right] \,,
	\label{eq:Vlong}
\end{align}

\begin{table}[b]
	\caption{
		Absolute neutrino masses (in eV) 
		for the extreme $m_\mathrm{min}$ values,
		and both normal and inverted hierarchies.
	}
	\label{tab:potential_masses}
	\begin{center}
		\begin{tabular}{ccc}
			\hline
			\hline
			\multicolumn{3}{c}{NH}\\
			\hline
			$m_3$ \hspace{0.2cm}	& 0.0500	&\hspace{0.2cm} 0.1118	\\
			$m_2$ \hspace{0.2cm}	& 0.0087	&\hspace{0.2cm} 0.1004	\\
			$m_1$ \hspace{0.2cm}	& 0			&\hspace{0.2cm} 0.1		\\
			\hline
			\hline
		\end{tabular}
		\hspace{0.5cm}
		\begin{tabular}{ccc}
			\hline
			\hline
			\multicolumn{3}{c}{IH}\\
			\hline
			$m_2$ \hspace{0.2cm}	& 0.0500	&\hspace{0.2cm} 0.1118	\\
			$m_1$ \hspace{0.2cm}	& 0.0492	&\hspace{0.2cm} 0.1115	\\
			$m_3$ \hspace{0.2cm}	& 0			&\hspace{0.2cm} 0.1		\\
			\hline
			\hline
		\end{tabular}
	\end{center}
\end{table}

\noindent
where 
$\mu^{}_{ij} = \frac{m_i m_j}{m_i + m_j}$ is the reduced mass of the 
$\nu_i \nu_j$ pair,
$\sqrt{t_{ij}^{}} = m_i + m_j$ 
and 
the \textcolor{DiracColor}{$1$}/\textcolor{MajoranaColor}{$0$} in the 
bracket 
correspond to \textcolor{DiracColor}{Dirac}/\textcolor{MajoranaColor}{Majorana} neutrinos.
The intermediate masses depend on the assumed 
either normal or inverted neutrino mass ordering 
and on the absolute mass $m_\mathrm{min}$ of the lightest neutrino. 
In Table~\ref{tab:potential_masses} 
we give the neutrino masses assuming 
$m_\mathrm{min} = 0$ and 
$m_\mathrm{min} = 0.1$~eV.

All the ingredients are now ready to compute at all distances the two-neutrino-mediated potential $V(r)$, 
always repulsive for aggregate matter, with the analytic limits given in Eqs.~(\ref{eq:Vshort}) and (\ref{eq:Vlong}).
We compare its $r$ dependence with the attractive gravitational potential. 
The six coherent weak charges are matter-dependent and their $Z$ behavior is different from the mass, 
opening the door to experimental studies based on the violation of WEP. 
The six $r$ dependencies are affected by the sought properties of 
absolute neutrino mass and the Dirac/Majorana nature. 
Each ($i,\, j$) channel has a different branching point $t_{ij} = (m_i + m_j)^2$, 
so the integral transforms of their absorptive parts have to be computed separately.
It is important to study whether the observable convoluted potential,
built from the six intermediate neutrino-pair exchanges,
still keeps near its range the precious information on the neutrino properties.

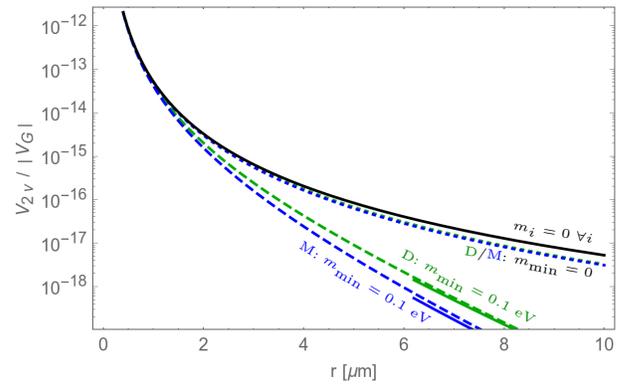
\begin{figure}
	\centering
	\begin{tikzpicture}[line width=1 pt, scale=1]
		\node at (0,0){
			\includegraphics[width=0.45\textwidth]{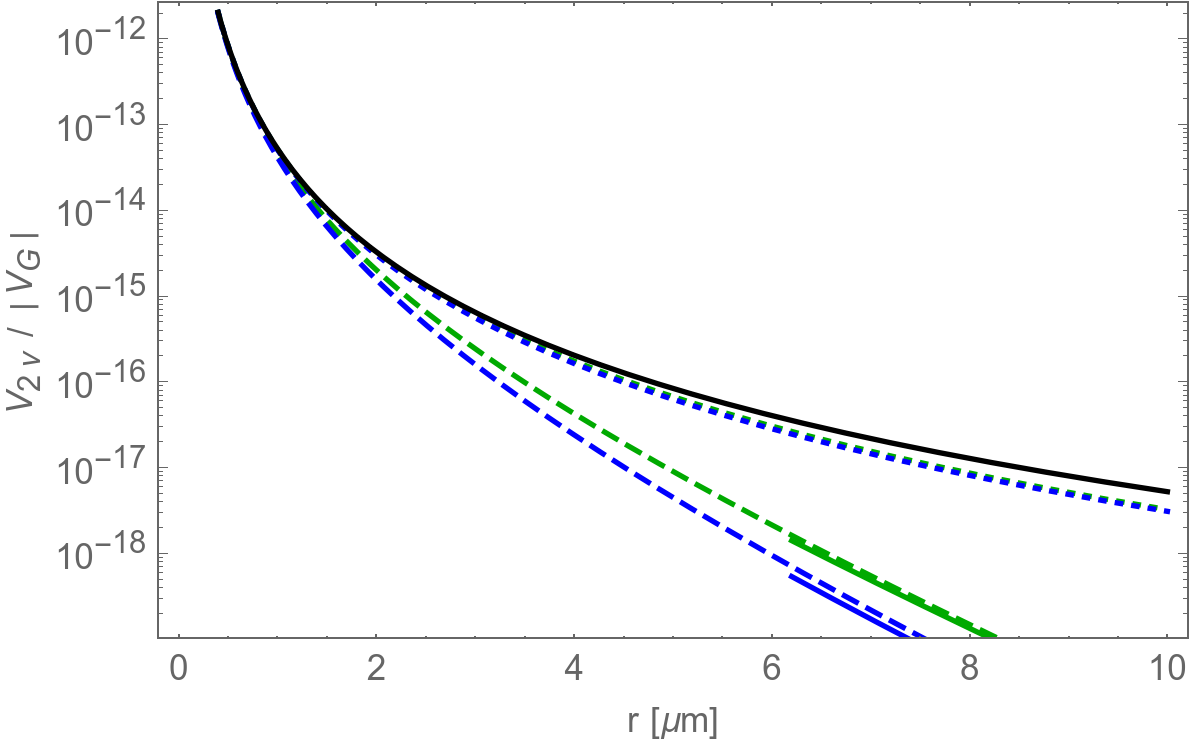}
		};

		\node[rotate=-10] at (3.8,-0.65)[left]{\tiny 
				$m_i = 0 \; \forall i$
		};

		\node[rotate=-10] at (3.8,-1.1)[left]{\tiny 
				\textcolor{DiracColor}{D}/\textcolor{MajoranaColor}{M}: $m_\mathrm{min}=0$
		};

		\node[rotate=-27] at (3.0,-1.77)[left]{\tiny 
				\textcolor{DiracColor}{D: $m_\mathrm{min}=0.1$~eV}
		};

		\node[rotate=-31] at (1.6,-1.8)[left]{\tiny 
				\textcolor{MajoranaColor}{M: $m_\mathrm{min}=0.1$~eV}
		};
	\end{tikzpicture}
	\caption[2nu-mediated long range potential]{
		2$\nu$-mediated long-range potential between two atoms of $^{56}$Fe,
		relative to their gravitational potential.
		Numerical results for $m_\mathrm{min} = 0$ 
		(superimposed \textcolor{DiracColor}{Dirac}/\textcolor{MajoranaColor}{Majorana} dotted lines)
		and $m_\mathrm{min}=0.1$~eV 
		(dashed \textcolor{DiracColor}{Dirac} and \textcolor{MajoranaColor}{Majorana} lines),
		together with analytic limits (solid lines) in Eqs.~(\ref{eq:Vshort}) and (\ref{eq:Vlong}).
	}
	\label{fig:potentials}
\end{figure}

We perform numerically the integrals
for the atom-atom interaction with the most stable nucleus, $^{56}$Fe,
and show the resulting 
Dirac and Majorana 2$\nu$-exchange potentials 
between 1 and 10 microns in Figure~\ref{fig:potentials}. 
This result demonstrates the existing major difference 
in that region of distances due to the value of $m_\mathrm{min}$
and the distinct $r$ dependence for Dirac and Majorana neutrino natures.
Indeed,
Eq.~(\ref{eq:Vlong}) explains the higher suppression of larger neutrino mass contributions,
as well as the S-wave Dirac potential being larger than the P-wave Majorana one.
On the other hand,
the two possible neutrino hierarchies, 
corresponding to the masses in Table~\ref{tab:potential_masses}, 
do not give an appreciable difference.

To summarize, 
a novel concept to search for the elusive neutrino properties 
of absolute mass and Dirac/Majorana nature has been presented. 
It is based on the existence of a coherent weak charge for neutral aggregate matter, 
with anticipated deviations of the weak equivalence principle, 
as well as the $r$ dependence of the two-neutrino-mediated potential at distances near its range. 
A novel methodology consisting in the (virtual) exploration of non-relativistic neutrinos 
with a different behavior depending on their massive nature is thus in place, 
complementary to the known approaches. 
The results in Figure~\ref{fig:potentials} for atom-atom interactions 
can be extended to sources of aggregate matter 
coherently enhancing both weak and gravitational potentials
while keeping the appreciable sensitivity to these fundamental neutrino properties.
It remains to be seen whether a terrestrial experiment could measure this weak observable
by means of devices able to cancel the gravitational effects of the Earth,
the major obstacle to overcome.
Current studies in this direction, looking for a conceptual design, are being pursued.

\begin{acknowledgments}
We would like to thank 
F.~Botella, A.~Magnon, S.~Palomares, D.~O.~Sabulsky, F.~S\'anchez and A.~Santamar\'ia for enlightening discussions. 
This research has been supported by 
the FEDER/MCIyU-AEI Grant FPA2017-84543-P, 
Severo Ochoa Excellence Centre Project SEV 2014-0398,
and
Generalitat Valenciana Projects GV PROMETEO 2017-033 and GV PROMETEO 2019-113.
One of us (A. S.) acknowledges the Spanish Ministry of Education (MECD) support through the FPU14/04678 grant.
\end{acknowledgments}

\bibliography{bibliography}

\end{document}